# Dynamic dipole and quadrupole phase transitions in the kinetic spin-1 model

Mustafa Keskin[a], Osman Canko[a], Ersin Kantar[b]
[a] Department of Physics, Erciyes University, 38039 Kayseri, Turkey
[b] Institute of Science, Erciyes University, 38039 Kayseri, Turkey

**Abstract**

The dynamic phase transitions have been studied, within a mean-field approach, in the kinetic spin-1 Ising model Hamiltonian with arbitrary bilinear and biquadratic pair interactions in the presence of a time varying (sinusoidal) magnetic field by using the Glauber-type stochastic dynamics. The nature (first- or second-order) of the transition is characterized by investigating the behavior of the thermal variation of the dynamic order parameters. The dynamic phase transitions (DPTs) are obtained and the phase diagrams are constructed in the temperature and magnetic field amplitude plane and found six fundamental types of phase diagrams. Phase diagrams exhibit one or two dynamic tricritical points depending on the biquadratic interaction (K). Besides the disordered (D) and ferromagnetic (F) phases, the FQ + D, F + FQ and F + D coexistence phase regions also exist in the system and the F and F + D phases disappear for high values of K.

## 1. Introduction

The spin-1 Ising model with arbitrary bilinear (J) and biquadratic (K) nearest-neighbor pair interactions, also known as the isotropic Blume-Emery-Griffiths (BEG) model, has been investigated theoretically [1] in connection with experimental results on magnetic phase transitions in some compounds [2]. It has subsequently been studied by the well known methods in equilibrium statistical physics such as, the mean-field approximation [3], the generalized constant coupling approximation [4], the effective field theory [5], the cluster variation methods [6] and the finite cluster approximation [7]. The exact solution of the model on the Bethe lattice was also studied [8].

There has also been much interest in understanding the nonequilibrium properties of the model. An early attempt to study the nonequilibrium behavior of the model was made by Obakata [9] who used the Bethe method and subsequently extended it into a time-dependent model, and obtained the relaxation times. Tanaka and Takahashi [10] studied the nonequilibrium behavior of the model within the conventional kinetic theory in the random-phase or generalized molecular-field approximation and obtained the relaxation curves of the order parameters. Keskin and co-workers [11, 12] have also studied a number of nonequilibrium behaviors of the model, in particular the "flatness" property of metastable states, the "overshooting" phenomenon and the phenomenon of frozen-in in metastable states as well as the role of the unstable states in the flow diagrams by using the path probability method (PPM) with point [11] and pair [12] distributions. Erdem and Keskin [13] studied the relaxation phenomena in the model near the phase transition temperatures within the Onsager's theory of irreversible thermodynamics. They also studied the critical behavior of the sound attenuation in the model, extensively [14]. Özer and Erdem [15] used the model to study dynamic of the voltage-gated ion channels in cell membranes by the PPM with point distribution.

On the other hand, some interesting problems in nonequilibrium systems are the nonequilibrium or the dynamic phase transition (DPT) and it is the one of the most important dynamic responses of current interests. The DPT was first found in a study within a mean-field approach the stationary states of the kinetic spin-1/2 Ising model under a time-dependent oscillating field [16, 17], by using the Glauber-type stochastic dynamics [18], and it was followed by Monte Carlo simulation, which allows the microscopic fluctuations, researches of kinetic spin-1/2 Ising models [19-23], as well as further mean-field studies [24]. Moreover, Tutu and Fujiwara [25] developed the systematic method for getting the phase diagrams in DPTs, and constructed the general theory of DPTs near the transition point based on mean-field description, such as Landau's general treatment of the equilibrium phase transitions. The DPT has also been found in a one-dimensional kinetic spin-1/2 Ising model with boundaries [26]. Reviews of earlier research on the DPT and related phenomena are found in Ref. 21. We should also mention that recent researches on the DPT are widely extended to more complex systems such as vector type order parameter systems, *e.g.*, the Heisenberg-spin systems [27], XY model [28], a Ziff-Gulari-Barshad model for CO oxidation with CO desorption to periodic variation of the CO pressure [29] and a high-spin Ising models such as the kinetic spin-1 BC model [30], the kinetic spin-3/2 BC model [31], the kinetic spin-3/2 BEG model [32], and a mixed-spin Ising model, *e.g.*, the kinetics of a mixed spin-1/2 and spin-1 Ising model [33]. Moreover, experimental evidences for the DPT has been found in ultrathin Co films on a Cu(001) surface [34] and in ferroic system (ferromagnets, ferroelectrics and ferroelastics) with pinned domain walls [35].

The present paper is aimed to study the dynamic phase transition (DPT) in the kinetic spin-1 Ising model Hamiltonian with arbitrary bilinear and biquadratic pair interactions in the presence of a time-dependent oscillating external magnetic field and construct the phase diagrams in the temperature and the magnetic-field amplitude plane. The time evolution of the system is described by the Glauber-type stochastic dynamics [18]. The nature (first- or second-order) of the transition is characterized by investigating the behavior of the thermal variation of the dynamic order parameters. We also calculate the Liapunov exponents to verify the stability of solutions and the DPT points

The paper is organized as follows. Section 2, the isotropic BEG model is presented briefly and the derivation of the mean-field (MF) dynamic equations of motion is given by using a Glauber-type stochastic dynamics in the presence of a time-dependent oscillating external magnetic field. Section 3, the stationary solutions of the dynamic equations are solved and the thermal behaviors of the dynamic order parameters are studied and as a result, the DPT points are calculated. Moreover, we also calculate the Liapunov exponents to verify the stability of solutions and the DPT points. Section 4, contains the presentation and the discussion of the phase diagrams. Finally, a summary is given in section 5.

## 2. The model and Derivation of Mean-Field Dynamic Equations of Motion

The Hamiltonian of spin-1 Ising model with arbitrary bilinear and biquadratic pair interactions, also called the isotropic BEG model, is given by

$$H = -J\sum_{<ij>} S_i S_j - K \sum_{<ij>} \left[ 3S_i^2 - 2 \right]\left[ 3S_j^2 - 2 \right] - H\sum_i S_i, \qquad (1)$$

where the spin located at site *i* on a discrete lattice can take values $\pm 1$ or 0 at each site *i* of a lattice and $\langle ij \rangle$ indicates a summation over all pairs of nearest-neighboring sites. J and K are, respectively, the nearest-neighbor bilinear and biquadratic exchange constants, and H is a time-dependent oscillating external magnetic field. H is given by

$$H(t) = H_0 \cos(wt), \tag{2}$$

where $H_0$ and $w = 2\pi v$ are the amplitude and the angular frequency of the oscillating field, respectively. The system is in contact with an isothermal heat bath at absolute temperature.

The order parameters of the system are the dipolar order parameter m, which is the excess of one orientation over the other orientation, also called the magnetization, and the quadrupolar order parameter q, that is a linear function of the average squared magnetization, given by

$$m \equiv \langle S_i \rangle, \tag{3}$$

and

$$q \equiv 3\langle S_i^2 \rangle - 2, \tag{4}$$

where $\langle ... \rangle$ is the thermal expectation value. The definition given by Eq. (4), which ensures that q=0 at infinite temperature, is different from the definition $q \equiv \langle S_i^2 \rangle$ which was used by Blume, Emery and Griffiths [36] and Lajzerowicz and Sivardière [37], Keskin and co-workers [38] and many other researchers.

Now, we apply the Glauber-type stochastic dynamics to obtain the mean-field dynamic equation of motion. Thus, the system evolves according to a Glauber-type stochastic process at a rate of $1/\tau$ transitions per unit time. We define $P(S_1, S_2, ..., S_N; t)$ as the probability that the system has the S-spin configuration, $S_1, S_2, ..., S_N$, at time t. The time-dependence of this probability function is assumed to be governed by the master equation which describes the interaction between spins and heat bath and can be written as

$$\frac{d}{dt} P(S_1, S_2, ..., S_N; t) = -\sum_i \left( \sum_{S_i \neq S_i'} W_i(S_i \to S_i') \right) P(S_1, S_2, ..., S_i, ..., S_N; t)$$
$$+ \sum_i \left( \sum_{S_i \neq S_i'} W_i(S_i' \to S_i) P(S_1, S_2, ..., S_i', ..., S_N; t) \right), \tag{5}$$

where $W_i(S_i \to S_i')$, the probability per unit time that the *i*th spin changes from the value $S_i$ to $S_i'$, and in this sense the Glauber model is stochastic. Since the system is in contact with a heat bath at absolute temperature T, each spin can change from the value $S_i$ to $S_i'$ with the probability per unit time;

$$W_i(S_i \to S_i') = \frac{1}{\tau} \frac{\exp(-\beta \Delta E(S_i \to S_i'))}{\sum_{S_i'} \exp(-\beta \Delta E(S_i \to S_i'))}, \tag{6}$$

where $\beta = 1/k_B T$, $k_B$ is the Boltzmann factor, $\sum_{S'_i}$ is the sum over the three possible values of $S'_i$, ±1, 0, and

$$\Delta E(S_i \to S'_i) = -(S'_i - S_i)(H + J\sum_{\langle j \rangle} S_j) - (S'^2_i - S^2_i)\left[3K\sum_{\langle j \rangle}(3S_j^2 - 2)\right] \quad (7)$$

gives the change in the energy of the system when the $S_i$-spin changes. The probabilities satisfy the detailed balance condition

$$\frac{W_i(S_i \to S'_i)}{W_i(S'_i \to S_i)} = \frac{P(S_1, S_2, \ldots, S'_i, \ldots, S_N)}{P(S_1, S_2, \ldots, S_i, \ldots, S_N)}, \quad (8)$$

and substituting the possible values of $S_i$, we get

$$W_i(1 \to 0) = W_i(-1 \to 0) = \frac{1}{\tau}\frac{\exp(-\beta y)}{2\cosh(\beta x) + \exp(-\beta y)}, \quad (9a)$$

$$W_i(1 \to -1) = W_i(0 \to -1) = \frac{1}{\tau}\frac{\exp(-\beta x)}{2\cosh(\beta x) + \exp(-\beta y)}, \quad (9b)$$

$$W_i(0 \to 1) = W_i(-1 \to 1) = \frac{1}{\tau}\frac{\exp(\beta x)}{2\cosh(\beta x) + \exp(-\beta y)}, \quad (9c)$$

where $x = H + J\sum_{\langle j \rangle} S_j$ and $y = 3K\sum_{\langle j \rangle}(3S_j^2 - 2)$. Notice that, since $W_i(S_i \to S'_i)$ does not depend on the value $S_i$, we can write $W_i(S_i \to S'_i) = W_i(S'_i)$, then the master equation becomes

$$\frac{d}{dt}P(S_1, S_2, \ldots, S_N; t) = -\sum_i \left(\sum_{S'_i \neq S_i} W_i(S'_i)\right) P(S_1, S_2, \ldots, S_i, \ldots, S_N; t)$$
$$+ \sum_i W_i(S_i)\left(\sum_{S'_i \neq S_i} P(S_1, S_2, \ldots, S'_i, \ldots, S_N; t)\right) \quad (10)$$

Since the sum of probabilities is normalized to one, by multiplying both sides of Eq. (10) by before $S_k$ then $(3S_k^2 - 2)$ and taking the average, we obtain

$$\tau\frac{d}{dt}\langle S_k \rangle = -\langle S_k \rangle + \left\langle \frac{2\sinh\beta(J\sum_j S_j + H)}{2\cosh\beta(J\sum_j S_j + H) + \exp(-3\beta K\sum_j(3S_j^2 - 2))} \right\rangle \quad (11)$$

$$\tau \frac{d}{dt}\langle 3S_k^2 - 2 \rangle = -\langle 3S_k^2 - 2 \rangle + 1 - \left\langle \frac{3\exp(-3\beta K \sum_j (3S_j^2-2))}{2\cosh \beta(J\sum_j S_j + H) + \exp(-3\beta K \sum_j (3S_j^2-2))} \right\rangle \quad (12)$$

These dynamic equations can be written in terms of a mean-field approach and hence the set of the mean-field dynamical equations of the system in the presence of a time-varying field are:

$$\tau \frac{d}{dt}\langle S \rangle = -\langle S \rangle + \frac{2\sinh \beta(Jz\langle S \rangle + H_0 \cos(wt))}{2\cosh \beta(Jz\langle S \rangle + H_0 \cos(wt)) + \exp(-3\beta Kz \langle 3S^2 - 2 \rangle)}, \quad (13)$$

$$\tau \frac{d}{dt}\langle 3S^2 - 2 \rangle = -\langle 3S^2 - 2 \rangle + 1 - \frac{3\exp(-3\beta Kz \langle 3S^2 - 2 \rangle)}{2\cosh \beta(Jz\langle S \rangle + H_0 \cos(wt)) + \exp(-3\beta Kz \langle 3S^2 - 2 \rangle)}, \quad (14)$$

where z is the coordination number. The system evolves according to the set of these coupled differential equations given by Eqs. (13) and (14) that can be written in the following form

$$\Omega \frac{d}{d\xi} m = -m + \frac{2\sinh [(1/T)(m + h\cos\xi)]}{2\cosh [(1/T)(m + h\cos\xi)] + \exp(-3kq/T)}, \quad (15)$$

$$\Omega \frac{dq}{d\xi} = -q + 1 - \frac{3\exp(-3kq/T)}{2\cosh [(1/T)(m + h\cos\xi)] + \exp(-3kq/T)}, \quad (16)$$

where $m \equiv \langle S \rangle$, $q \equiv 3\langle S^2 \rangle - 2$, $\xi = wt$, $T = (\beta zJ)^{-1}$, $k = \frac{K}{J}$, $h = H_0/zJ$, and $\Omega = \tau w$. Hence, the set of the mean-field dynamical equations for the order parameters are obtained. We fixed z=4 and $\Omega = 2\pi$. Solution and discussion of these equations are given in the next section.

### 3. Thermal Behaviors of Dynamic Order Parameters and Dynamic Phase Transition Points

In this section, we shall first solve the set of dynamic equations and present the behaviors of average order parameters in a period as a function of the reduced temperature and as a result, the DPT points are calculated. Moreover, we also calculate the Liapunov exponent to verify the stability of solutions and the DPT points. For these purposes, first we have to study the stationary solutions of the set of dynamic equations, given in Eqs. (15) and (16), when the parameters T, k and h are varied. The stationary solutions of Eqs. (15) and (16) will be a periodic function of $\xi$ with period $2\pi$; that is, $m(\xi + 2\pi) = m(\xi)$ and $q(\xi + 2\pi) = q(\xi)$. Moreover, they can be one of three types according to whether they have or do not have the property

$$m(\xi + \pi) = -m(\xi), \quad (17a)$$

and

$$q(\xi+\pi) = -q(\xi). \tag{17b}$$

A solution satisfies both Eqs. (17a) and (17b) are called a symmetric solution which corresponds to a disordered (D) solution. In this solution, the magnetization $m(\xi)$ always oscillates around the zero value and is delayed with respect to the external magnetic field. On the other hand, the quadrupolar order parameters $q(\xi)$ oscillates around a non zero value for finite temperature and around a zero value for infinite temperature due to the reason that q=0 at infinite temperature by the definition of q, given in Eq. (4). The second type of solution, which does not satisfy Eqs. (17a) and (17b), is called a nonsymmetric solution that corresponds to a ferromagnetic (F) solution. In this case the magnetization and quadrupolar order parameters do not follow the external magnetic field any more, but instead of oscillating around a zero value; they oscillate around a nonzero value. The third type of solution, which satisfies Eq. (17a) but does not satisfy Eq. (17b), corresponds to ferroquadrupolar or simply quadrupolar (FQ) phase. In this solution, $m(\xi)$ oscillates around the zero value and are delayed with respect to the external magnetic field and $q(\xi)$ does not follow the external magnetic field any more, but instead of oscillating around a zero value, it oscillates around a nonzero value, namely either -2 or +1. If it oscillates around -2, this nonsymmetric solution corresponds to the ferroquadrupolar or simply quadrupolar (FQ) phase and if it oscillates around +1, this corresponds to the disorder phase (D); because this solution does not give a phase transition, this fact is seen explicitly in Fig. 2(e). Moreover, for high values of temperature this solution corresponds to the symmetric solution, hence for the infinite temperature, Fig. 1(c) becomes Fig. 1(a) except $q(\xi)$ oscillates around -2. These facts are seen explicitly by solving Eqs. (15) and (16) numerically. Eqs. (15) and (16) are solved by using the numerical method of the Adams-Moulton predictor corrector method for a given set of parameters and initial values and presented in Fig. 1. From Fig. 1, one can see five different solutions, namely the D, F phases or solutions and three coexistence solutions, namely the FQ + D in which FQ, D solutions coexist, the F + FQ in which F, FQ solutions coexist and F + D in which F, D solution coexist, exist in the system. In Fig. 1(a) only the symmetric solution is always obtained, hence we have a disordered (D) solution, but in Fig. 1(b) only the nonsymmetric solution is found; therefore, we have a ferromagnetic (F) solution. Neither solutions depends on initial values. In Fig. 1 (c), we have a sysmetric solution for $m(\xi)$ and the nonsymmetric solution for $q(\xi)$, because $m(\xi)$ oscillate around zero values and $q(\xi)$ around -2 or +1. As explained above, the solution of $q(\xi)$ which oscillates around +1 does not give a phase transition, see Fig. 2(e) and it corresponds to the D phase, hence we have the coexistence solution (FQ + D). On the other hand, in Fig. 1(d) we have two solutions for both $m(\xi)$ and $q(\xi)$. The first solution, $m(\xi)$ oscillates around zero and $q(\xi)$ around -2, hence we have FQ phase and the second one $m(\xi)$ oscillates around $\pm 1$ and $q(\xi)$ around +1, thus we have F phase. Therefore, in this case the F + FQ coexistence region occurs in the system and the solutions depend on the initial values, seen in Fig. 1(d) explicitly. Fig. 1(e) is similar to the Fig. 1(d), except F and D phases exist in Fig. 1(e). Hence, we have F + D coexistence solution and these solutions also depend on the initial values.

Thus, Fig. 1 displays that we have five phases in the system, namely D, F, FQ + D, F + FQ and F + D solutions or phases. In order to see the dynamic boundaries among these five phases, we have to calculate DPT points and then we can present phase diagrams of the system. DPT points will be obtained by investigating the behavior the average order parameters in a period or the dynamic order parameters as a function of the reduced temperature. These investigations will be also checked and verified by calculating the Liapunov exponents.

The dynamic order parameters, namely the dynamic magnetization (M) and the dynamic quadrupole moment (Q), are defined as

$$M = \frac{1}{2\pi} \int_0^{2\pi} m(\xi) d\xi, \tag{18}$$

$$Q = \frac{1}{2\pi} \int_0^{2\pi} q(\xi) d\xi. \tag{19}$$

The behavior of M and Q as a function of the reduced temperature for several values of h and k are obtained by combining the numerical methods of Adams-Moulton predictor corrector with the Romberg integration and the results are plotted in Figs. 2(a)-(e). In these figures, $T_C$ and $T_t$ are the critical or the second-order phase transition and first-order phase transition temperatures for both M and Q, respectively and $T_{tQ}$ is the first-order phase transition temperatures for only Q. Fig. 2(a) represents the reduced temperature dependence of the dynamic order parameters, M and Q, for k=0.1 and h=0.6. In this case, M decreases to zero continuously as the reduced temperature increases, therefore a second-order phase transition occurs. On the other hand, Q decreases until $T_C$, as the temperature increases, and at $T_C$, it makes a cusp and then decreases to zero as the temperature increases. In this case the phase transition is from F phase to D phase. Figs. 2(b) and 2(c) illustrate the thermal variations of M and Q for k=0.1 and h=0.2 for two different initial values; i.e., the initial values of M and Q are taken one for Fig. 2(b) and M=0 and Q=-2 for Fig. 2(c). The behavior of Fig. 2(b) is similar to Fig. 2(a), hence the system undergoes a second-order phase transition. In Fig. 2(c), the system undergoes two successive phase transitions, the first one is a first-order from the quadrupolar (FQ) phase to the ferromagnetic (F) phase and the second one is a second-order, from the F phase to the D phase. This means that the coexistence region, i.e., the F + FQ phase, exists in the system and this fact is seen in the phase diagram of Fig. 4(a) explicitly, compare in Figs. 2(b) and 2(c) with Fig. 4(a). Finally, Figs. 2(d) and (e) show the behavior of M and Q as a function of the reduced temperature for k=0.1 and h=0.8 for two different initial values; i.e., the initial value of M and Q are taken one for Fig. 2(d) and M=0 and Q=-2 for Fig. 2(e). In Fig. 2(d), both M and Q undergo a first-order phase transition, because M and Q decrease to zero discontinuously as the reduced increases and the phase transition is from the F phase to the D phase. Fig. 1(e) shows that M always equals to zero and Q=1 at zero temperature but does not undergo any phase transition. This implies that the nonsymmetric solution of $q(\xi)$ that oscillates around +1, does not undergo phase transition. Hence this figure corresponds to the D phase.

Now we can check and verify the stability of solutions, and as well as the DPT points by calculating the Liapunov exponent. If we write Eqs. (15) and (16) as

$$\Omega \frac{dm}{d\xi} = F_1(m, \xi), \tag{20}$$

$$\Omega \frac{dq}{d\xi} = F_2(q, \xi), \tag{21}$$

then the Liapunov exponents $\lambda_m$ and $\lambda_q$ are given by

$$\Omega \lambda_m = \frac{1}{2\pi} \int_0^{2\pi} \frac{\partial F_1}{\partial m} d\xi, \tag{22}$$

$$\Omega \lambda_q = \frac{1}{2\pi} \int_0^{2\pi} \frac{\partial F_2}{\partial q} d\xi. \tag{23}$$

When $\lambda_m < 0$ and $\lambda_q < 0$, the solution is stable. We have two Liapunov exponents, namely, one is associated to the symmetric solution, $\lambda_{ms}$ and $\lambda_{qs}$, and the other to the nonsymmetric solution, $\lambda_{mn}$ and $\lambda_{qn}$, for both m and q. If $\lambda_{ms}$ and $\lambda_{mn}$ increase to zero continuously as the reduced temperature approaches to the phase transition temperature, the temperature where $\lambda_{mn} = \lambda_{ms} = 0$ is the second-order phase transition temperature, $T_C$. Moreover, if $\lambda_{qn}$ and $\lambda_{qs}$ increase continuously as the reduced temperature approaches to the phase transition temperature and then the temperature where $\lambda_{qn}$ and $\lambda_{qs}$ make a cusp is the second-order phase transition temperature, $T_C$. The reason $\lambda_{qn}$ and $\lambda_{qs}$ are not zero at $T_C$ due to the cause that Q is not zero at $T_C$ and it is zero at infinite temperature. On the other hand, if the Liapunov exponent approaches the phase transition temperature, the temperature at which the Liapunov exponents make a jump discontinuity is the first-order phase transition temperature. In order to see these behaviors explicitly, the values of the Liapunov exponents are calculated and plotted as a function the reduced temperature for k=0.1 and h=0.2 (these values correspond to Fig. 2(c)), seen in Fig. 3. In the figure thick and thin lines represent the $\lambda_s$ and $\lambda_n$, respectively, and $T_C$ is the second-order phase transition temperature for M and Q and $T_{tQ}$ is the first-order phase transition temperature for only Q. In Fig. 3, the system undergoes successive phase transitions: First, the phase transition is a first-order, because of $\lambda_{ms}$ and $\lambda_{mn}$ make a jump discontinuity at $T_{tQ}$=0.1775, the second one is a second-order phase transition, because $\lambda_{mn} = \lambda_{ms} = 0$ at $T_C$=0.6525, seen in Fig. 3(a). Fig. 3(b) illustrates the behavior of Liapunov exponents for q. It is seen from this figure that, first both $\lambda_{qn}$ and $\lambda_{qn'}$ make a jump discontinuity, hence we have a first-order phase transition at $T_{tQ}$=0.1775 ($\lambda_{qn'}$ corresponds to the FQ phase and $\lambda_{qn}$ corresponds to the F phase); then $\lambda_{qn}$ and $\lambda_{qs}$ make a cusp, hence the second-order phase transition temperature occurs at $T_C$=0.6525. If one compares Fig. 3 with Fig. 2(c) one can see that $T_{tQ}$ and $T_C$ found by using the both calculations are exactly the same. Moreover, we have also verified the stability of the solution by this calculation, because we have always found that $\lambda_m$ <0 and $\lambda_q$ <0.

Finally, it is worthwhile to mention that the oscillating external magnetic field induces the phase transition, because if one has done the calculations for the reduced external magnetic field amplitude h, one can see that the system does not undergo any phase transitions. This fact was illustrated in Fig. 6 of Ref. 30.

### 4. Phase Diagrams

Since we have obtained and verified the DPT points in Section 3, we can now present the phase diagrams of the system. The calculated phase diagrams in the (T, h) plane are presented in Fig. 4 for various values of k. In these phase diagrams, the solid and dashed lines represent the second- and first-order phase transition lines, respectively. The dynamic tricritical point is denoted by a filled circle. As seen from the figure, the following six main topological different types of phase diagrams are found.

(i) For 0<k≤0.111, Fig. 4(a) represents the phase diagram in the (T,h) plane for k=0.1. In this phase diagram, at high reduced temperature (T) and reduced external magnetic field amplitudes (h) the solutions are disordered (D) and at low values of h and high values of T, they are ferromagnetic (F). The boundary between these regions, F→D, is the second-order phase line. At low reduced temperatures, there is a range of values of h in which the F and D phases coexist, called the coexistence region or phase, F + D. The F + D region is separated from the F and the D phases by the first-order phase line. The system also exhibits only one dynamic tricritical point where the both first-order phase transition lines merge and signals the change from a first- to a second-order phase transitions. Moreover, very low T and h values one more F+FQ coexistence region also exists and the dynamic phase boundaries among these coexistences phases and the F phase, and between the F + D phase and the D phase are all first-order lines.

(ii) For 0.111<k≤0.169, the phase diagram is displayed for k=0.15 and shown in Fig. 4(b). The phase diagram is similar to Fig. 4(a), except the F phase disappears for very low values of T as well as at zero temperature and the F + FQ coexistence region becomes large, seen in the figure.

(iii) For 0.169<k≤0.293, we are performed the phase diagram at k=0.2, seen in Fig. 4(c). In this type the system exhibits two dynamic tricritical points that one of them occurs similar place as in Fig. 4(a) and the other occurs in the low values of h and high values of T, thus for the very low values of h and high values of T, the dynamic phase boundary between the F and D phase is a first-order phase line. For low values of T, there is a range of values of h in which the FQ + D phase occurs, seen in Fig. 4(c). Moreover, the F + D phase also occurs in the system for high values of T and low values of h. The dynamic phase boundaries among the these five different phases are first-order lines, except the boundary connecting the two dynamic tricritical points that separates the F phase from the D phase, this boundary is a second-order line.

(iv) For 0.293<k≤0.325, in this type the phase diagram is presented for k=0.3, seen in Fig. 4(d) and is similar to the type (iii), except that the FQ + D region becomes large and the F+D phase appears at low values h and high values of T.

(v) For 0.325<k≤0.535, the phase diagram is obtained for k=0.4, seen in Fig. 4(e), and three phases, namely the F + FQ, FQ + D and D phases, exist. For low values of T and h, the F+FQ phase occurs and the dynamic phase boundary between the F + FQ and FQ + D phases is a first-order line for low values of T and high values of h and also high values of T and very low values of h; hence the boundary between these two first-order lines is a second-order line, seen in the figure. Therefore, the system exhibits two dynamic tricritical points. On the other hand, the dynamic phase boundary between the FQ + D and D phases is a first-order phase line.

(vi) For k>0.535, the phase diagram is constructed for k=1.0 and is similar to the type (v), except one of the dynamic tricritical point, that occurs at low values of h, disappears, hence the system exhibits only one dynamic tricritical point, seen in Fig. 4(f). Moreover, if one increases k values, the regions of the F + FQ and FQ + D phases also become greater.

## 5. Summary

We have studied within a mean-field approach the stationary states of the kinetic spin-1 Ising model Hamiltonian with arbitrary bilinear and biquadratic pair interactions, also called the isotropic Blume-Emery-Griffiths (BEG) model, in the presence of a time-dependent oscillating external magnetic field. We use a Glauber-type stochastic dynamics to describe the time evolution of the system. We have studied the behavior of the time-dependence of the order parameters, namely magnetization or the dipole moment and the quadruple moment, and the behavior of the average order parameters in a period as a function of reduced temperature, and found that the behavior of the system strongly depends on the biquadratic pair interaction. The DPT points are obtained and the phase diagrams presented in the (T, h) plane and six different phase diagrams are found. The system exhibits the D, F phases and/or the FQ + D, F + FQ, F + D coexistence regions depending on k values and the dynamic phase boundaries among these phases and coexistence regions are first-order lines in general and a second-order for between the F and D phases, and between the F + FQ and FQ + D phases in few cases. Therefore, one or two dynamic ticritical points also occur. Finally, we should also mention that we have also calculated the Liapunov exponents to verify the stability of solutions and the DPT points.

Finally, it is worthwhile to mention that there is a strong possibility that at least some of the first-order transitions and multicritical points seen in the mean-field results are very likely artifacts of the approximation. This is because, for field amplitude less than the coercive field (at the temperature less than the static ferro-para (or order-disorder) transition temperature), the response magnetization varies periodically but asymmetrically even in the zero-frequency limit; the system then remains locked to the higher, yet locally attractive, well of the free energy and can not go the other well, in the absence of noise or fluctuations [19, 20(d), 21(a), 24(a and c), 39]. However, this mean-field dynamic study suggests that the spin-1 BEG model Hamiltonian with arbitrary bilinear and biquadratic pair interactions in the presence of a time dependent oscillating external magnetic field has an interesting dynamic behavior, quite different from the standard Ising model. Therefore, it would be worthwhile to further study it with more accurate techniques such as dynamic Monte Carlo simulations or renormalization group calculations

### Acknowledgements
This work was supported by the Technical Research Council of Turkey (TÜBİTAK) Grant No. 105T114 and Erciyes University Research Funds Grant No: FBA-06-01. We are very grateful to Bayram Deviren for useful discussions.

# List of the Figure Captions

**Fig. 1.** Time variations of the magnetization (m) and the quadrupolar order parameter (q):
  a) Exhibiting a disordered phase (D), k=0.1, h=1.0 and T=2.0
  b) Exhibiting a ferromagnetic phase (F), k=0.1, h=0.2 and T=0.375
  c) Exhibiting a coexistence region (FQ+D), k=0.3, h=1.4 and T=0.125.
  d) Exhibiting a coexistence region (F+FQ), k=0.3, h=0.3 and T=0.25.
  e) Exhibiting a coexistence region (F+D), k=0.1, h=0.75 and T=0.175.

**Fig. 2.** The reduced temperature dependence of the dynamic magnetization (M) and (the thick solid line) and the dynamic quadruple moment (Q) (the thin solid line). $T_C$ and $T_t$ are the critical or the second-order phase transition and the first-order phase transition temperature for both M and Q, respectively and $T_{tQ}$ is the first-order phase transition temperature for only Q.

   a) Exhibiting a second-order phase transition from the F phase to the D phase for k=0.1 and h=0.6; 0.5125 is found $T_C$.

   b) Exhibiting a second-order phase transition from the F phase to the D phase for k=0.1 and h=0.2; 0.6525 is found $T_C$.

   c) Exhibiting two successive phase transitions, the first one is a first-order phase transition from the FQ phase to the F phase and the second one is second-order phase transition the F phase to the D phase for k=0.1 and h=0.2; 0.6525 and 0.1775 found $T_C$ and $T_{tQ}$, respectively.

   d) Exhibiting a first-order phase transition from the F phase to the D phase for k=0.1 and h=0.8; 0.2125 is found $T_t$.

   e) The system does not undergo any phase transition and corresponds to the D phase; k=0.1 and h=0.8.

**Fig. 3.** The values of the Liapunov exponents as a function the reduced temperature (T) for k=0.1 and h=0.2. Thick and thin lines represent the $\lambda_s$ and $\lambda_n$, $\lambda_{n'}$, respectively, and $T_C$ are the critical or the second-order phase transition for both M and Q, respectively and $T_{tQ}$ is the first-order phase transition temperatures for only Q.

   **a)** The behavior of the Liapunov exponents as a function of T for m. The system undergoes two successive phase transitions: First, the phase transition is a first-order, because $\lambda_{ms}$ and $\lambda_{mn}$ make a jump discontinuity and the first-order transition occurs at $T_{tQ}$=0.1775; the second one is a second-order phase transition, because $\lambda_{mn} = \lambda_{ms} = 0$ at $T_C$=0.6525.

   **b)** Same as (a), but for q. Both $\lambda_{qn}$ and $\lambda_{qn'}$ make a jump discontinuity, hence we have a first-order phase transition at 0.1775 ($\lambda_{qn'}$ corresponds to the FQ phase and $\lambda_{qn}$ corresponds to the F phase); then $\lambda_{qn}$ and $\lambda_{qs}$ make a cusp, hence the second-order phase transition temperature occurs at $T_C$=0.6525.

**Fig. 4.** Phase diagrams of the isotropic Blume-Emery-Griffiths model in the (T, h) plane. The disordered (D), ferromagnetic (F) and three different the coexistence phase regions, namely the FQ+D, F+FQ and F+D regions, are found. Dashed and solid lines represent the first- and second-order phase transitions, respectively, and the dynamic tricritical points are indicated with filled circles. **a)** k=0.1, **b)** k=0.15, **c)** k=0.2, **d)** k=0.3, **e)** k=0.4 and **f)** k=1.0.

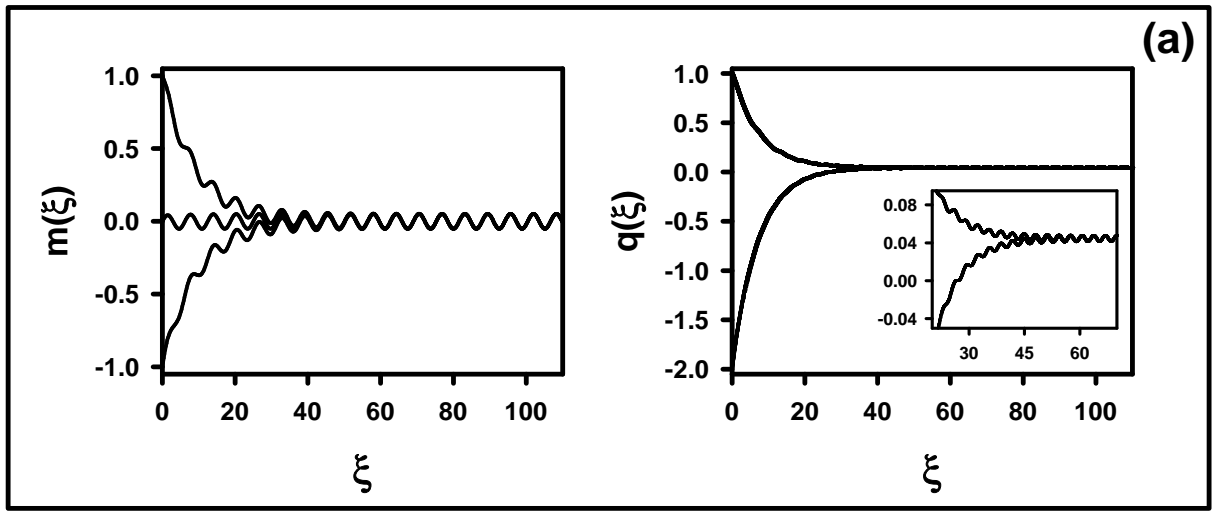

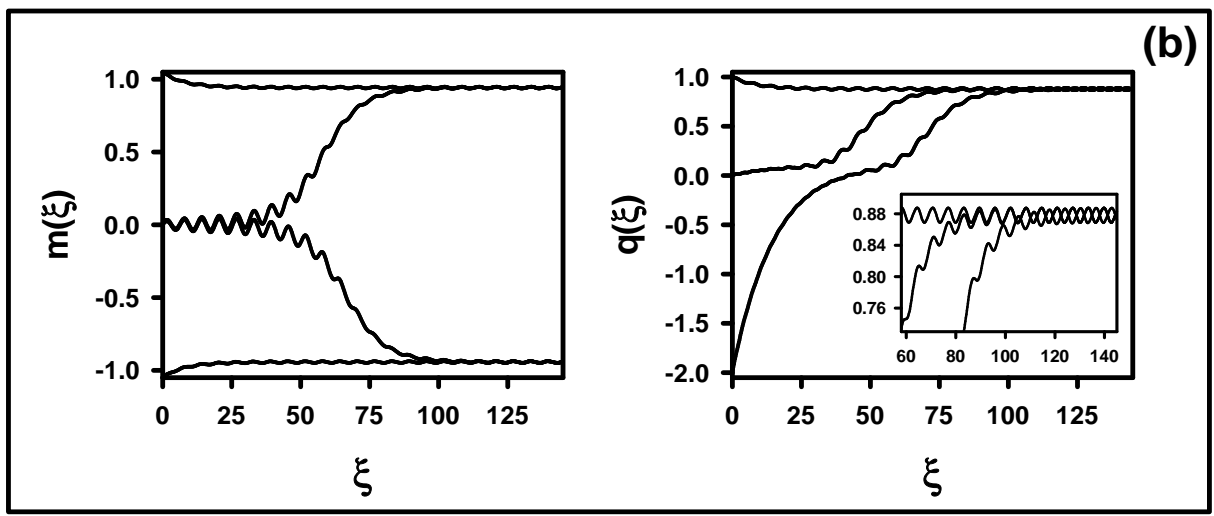

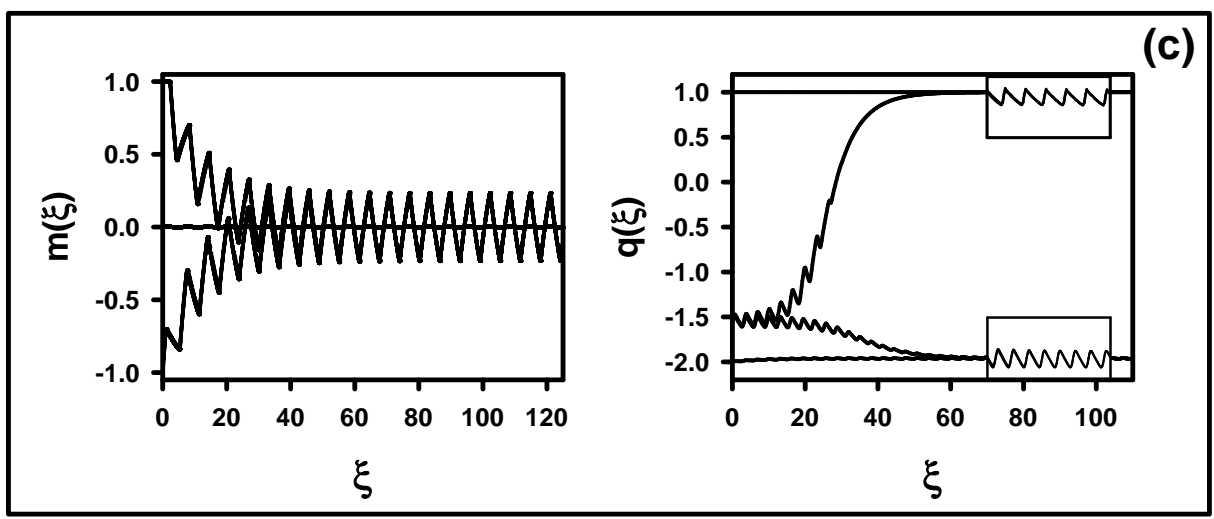

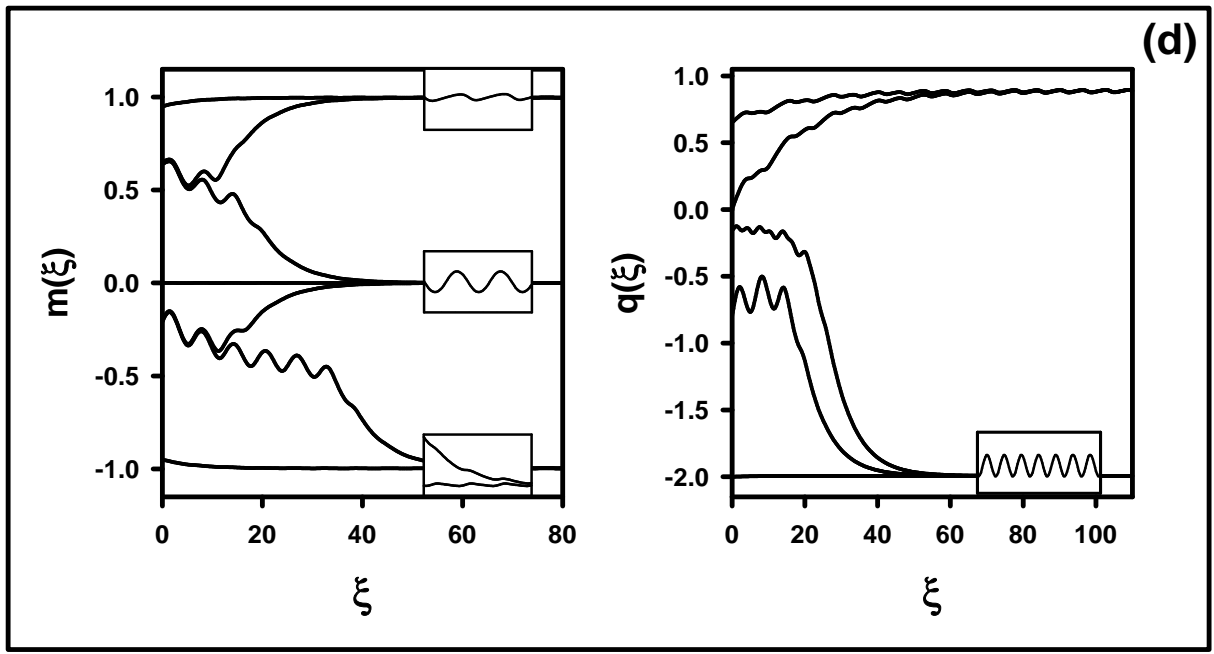
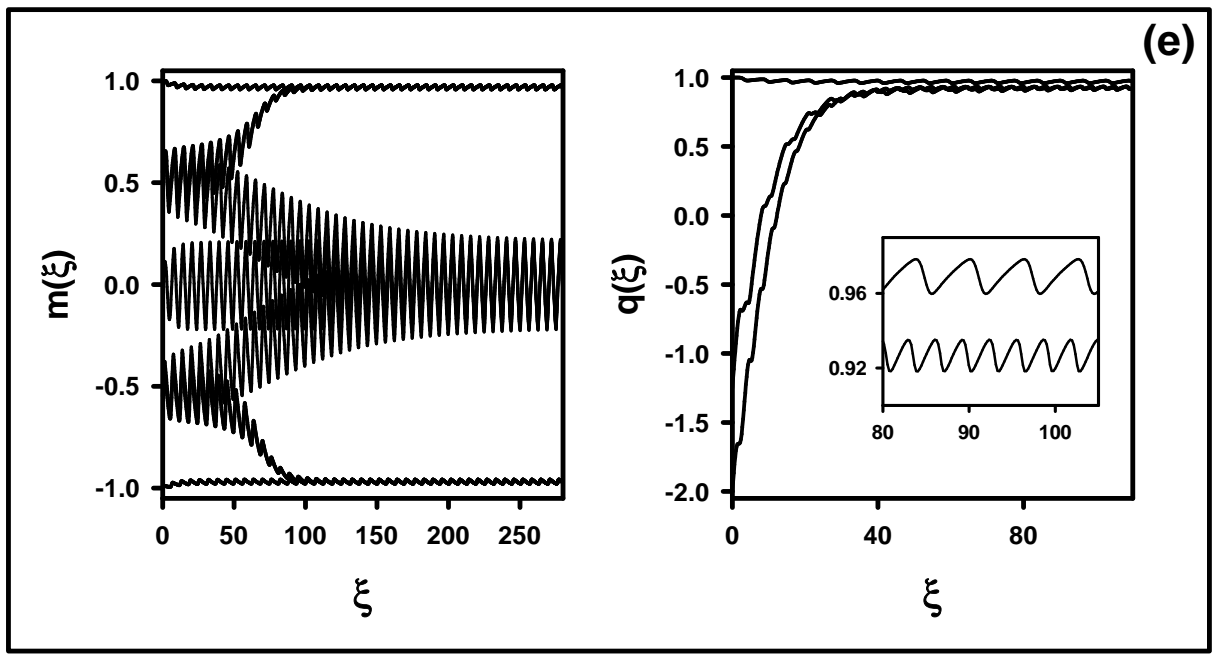

Fig.1

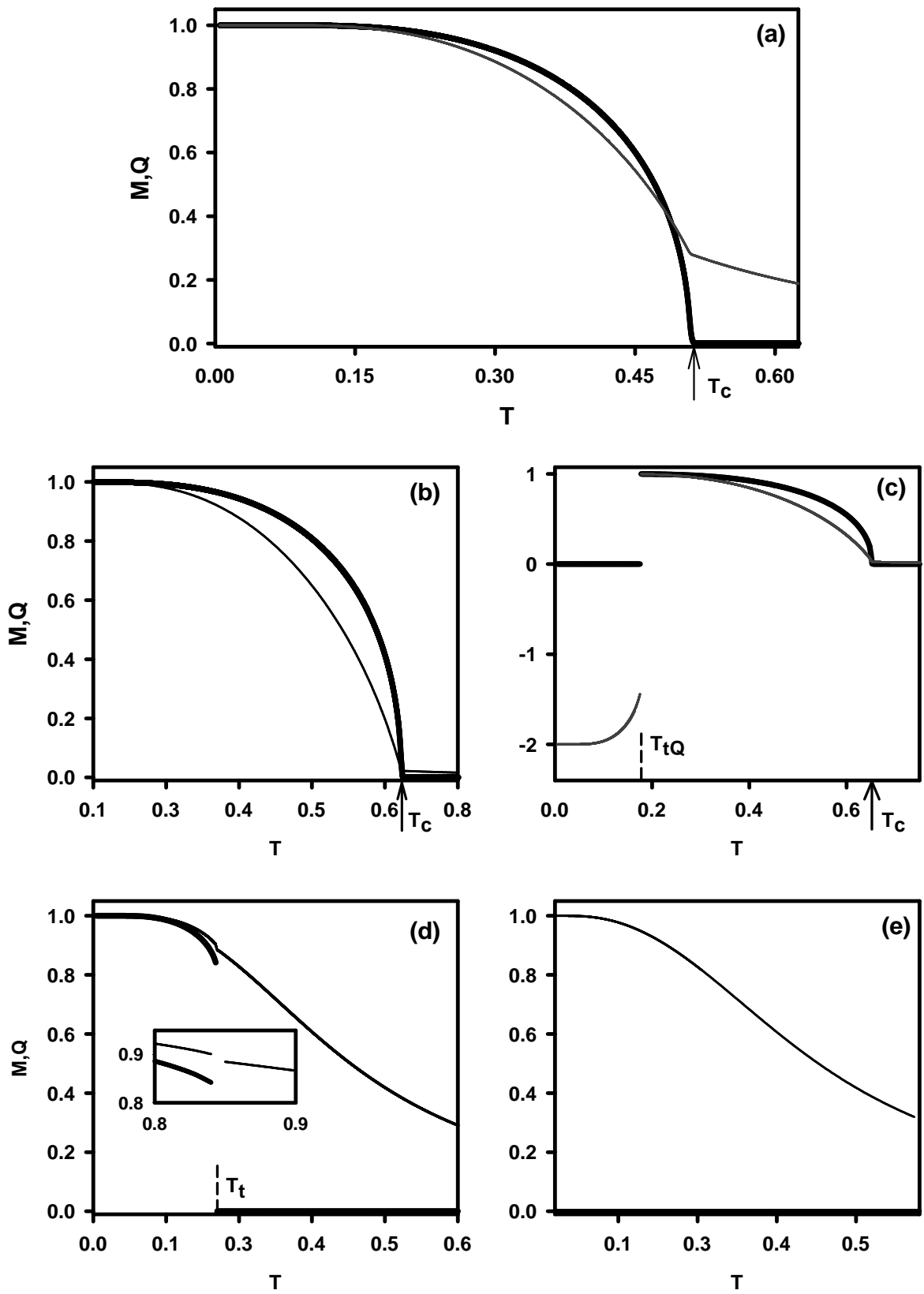

Fig. 2

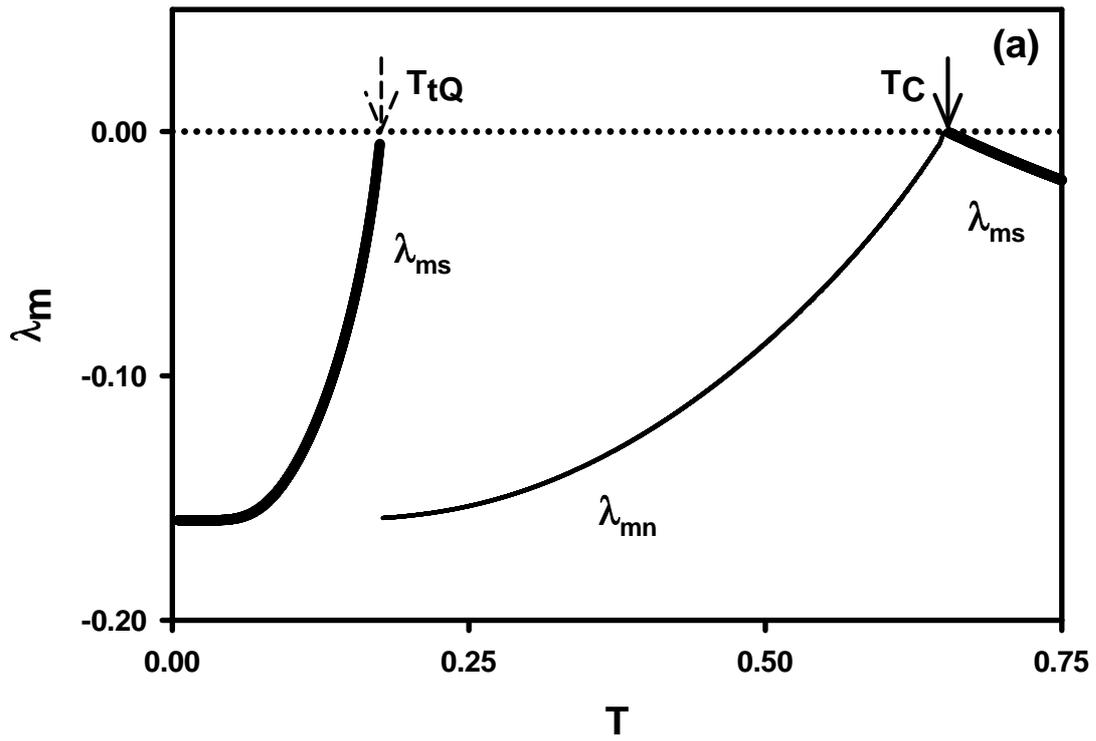
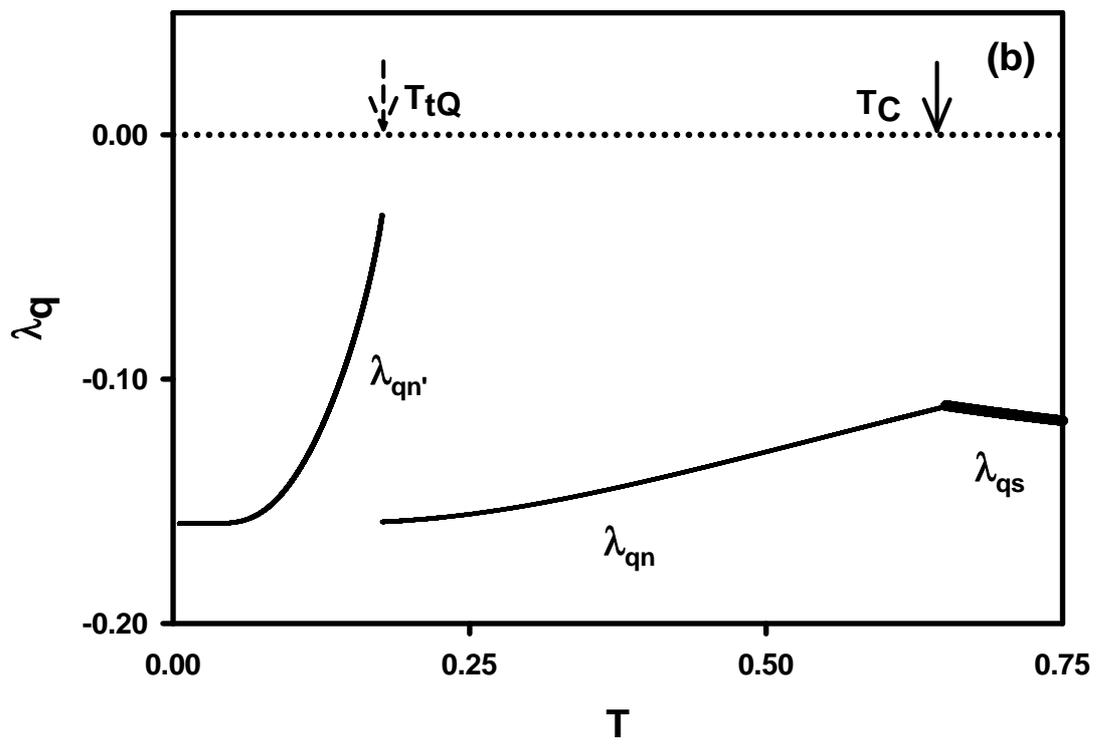

Fig. 3

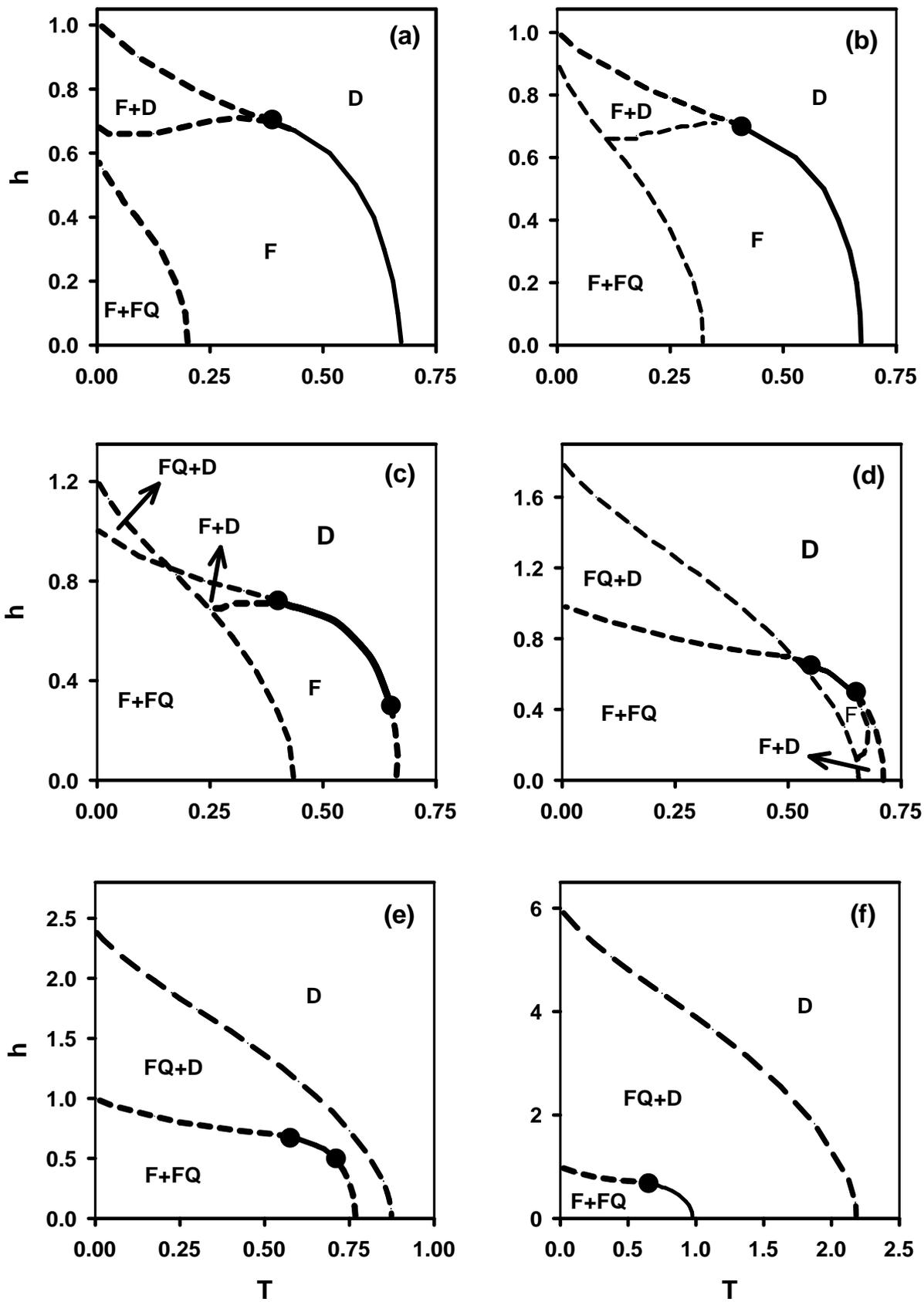

Fig. 4